\documentclass[english,twoside,a4paper,10pt]{article}

\usepackage[latin1]{inputenc}
\usepackage[T1]{fontenc}
\usepackage{amsmath}
\usepackage{amsfonts}
\usepackage{graphicx}
\usepackage{subfigure}
\usepackage{a4wide}
\usepackage{amssymb}
\usepackage{fancyhdr}
\usepackage{mathrsfs}

\begin{document}

\title{\bf Nearly Starobinsky inflation from modified gravity}
\author{L. Sebastiani$^{1}$\footnote{Author to whom correspondence should be 
addressed. E-mail:l.sebastiani@science.unitn.it}\,,
G. Cognola $^2$\footnote{E-mail:cognola@science.unitn.it}\,,
R. Myrzakulov$^1$\footnote{Email: rmyrzakulov@gmail.com; 
rmyrzakulov@csufresno.edu}\,
S.~D. Odintsov$^{3,4,5}$\footnote{E-mail: odintsov@ieec.uab.es}
and S. Zerbini$^2$\footnote{E-mail address:zerbini@science.unitn.it}\\
\\
\begin{small}
$^1$ Eurasian International Center for Theoretical Physics and  Department of 
General
\end{small}\\
\begin{small}
Theoretical Physics, Eurasian National University, Astana 010008, Kazakhstan
\end{small}
\\
\begin{small}
$^2$ Dipartimento di Fisica, Universit\`a di Trento, Italy and
\end{small}\\
\begin{small}
  Gruppo Collegato di Trento, Istituto Nazionale di Fisica Nucleare, Sezione di 
Padova, Italy
\end{small}\\
\begin{small}
$^3$Consejo Superior de Investigaciones Cient\'{\i}ficas, ICE/CSIC-IEEC,
Campus UAB,
\end{small}\\
\begin{small}
  Facultat de Ci\`{e}ncies, Torre C5-Parell-2a pl, E-08193
Bellaterra (Barcelona), Spain
\end{small}\\
\begin{small}
$^4$Instituci\'{o} Catalana de Recerca i Estudis Avan\c{c}ats
(ICREA), Barcelona, Spain
\end{small}\\
\begin{small}
$^5$ Tomsk State Pedagogical University, 634061,Tomsk, Russia
\end{small}
}

\date{}

\maketitle


\begin{abstract}

We study inflation induced by (power-low) scalar curvature corrections to General Relativity. The class of inflationary scalar potentials $V(\sigma)\sim\exp[n\,\sigma]$, $n$ general parameter, is investigated in 
the Einsein frame and the corresponding actions in the Jordan frame are 
derived. We found the conditions for which these potentials are able to reproduce viable inflation according with the last cosmological data and lead to 
large scalar curvature corrections which emerge only at a mass scale larger than the 
Planck mass. Cosmological constant may appear or be set equal to zero in the Jordan frame action without changing the behaviour of the model during inflation.
Moreover, polynomial corrections to General Relativity are analyzed in detail. 
When  de Sitter space-time emerges as an exact solution of the models, it is 
necessary to use perturbative equations in the Jordan framework to study their 
dynamics during the inflation. In this case, we demonstrate that 
the Ricci scalar decreases after a correct amount of inflation, making the 
models consistent with the observable evolution of the universe.
 \end{abstract}



\section{Introduction}

Large number of inflationary cosmology models is based on scalar fields, which 
play an important role in the particle physics theories. The inflation is 
produced by an homogeneous scalar field, dubbed  inflaton, which under suitable 
conditions may lead to an early-time accelerated expansion. Following the first 
proposal of Guth~\cite{Guth} and Sato~\cite{Sato}, in the last years many 
inflationary models based on scalar fields ( and inspired by modified gravity 
theories, string theories, quantum effects in the hot universe, etc) have been 
proposed.

Typically the magnitude of scalar field is very large at the beginning of the 
inflation and
then it rolls down towards a potential minimum where the inflation ends (see 
Ref.~\cite{chaotic} as an example of chaotic inflation). In other models the 
field can fall in a potential hole where it starts to oscillate and the 
rehating processes take place~\cite{buca1, buca2, buca3, buca4}.
Some more complicated models are based on a phase transition between two scalar 
fields: they are the so-called hybrid or double inflation models~\cite{ibrida1, 
ibrida2}.
For the introduction to the dynamics of inflation see Ref.~\cite{Linde} and 
Refs.~\cite{revinflazione}--\cite{rr2}.

Recently, cosmological and astrophysical data~\cite{cosmdata} seem to confirm 
the predictions of Starobinsky inflationary model~\cite{Staro}. Such a model is 
based on the account of $R^2$-term as the correction in the Einstein equations. 
This quadratic correction emerges in the Planck epoch and plays a fundamental 
role in the high curvature limit, when the early-time acceleration takes place. 
Such theory is conformally equivalent to a scalar-tensor theory in the Einstein 
frame, where the inflaton drives the expansion in a quasi-de Sitter space-time 
and slowly moves to the end of inflation, when the reheating 
processes~\cite{r1,r2,r3} start.
Such  inflationary model has been recently revisited in many works. Among them, 
in Ref.~\cite{genStar} a superconformal generalization of such a model in 
superconformal theory has been investigated, and in 
Refs.~\cite{SS1}--\cite{SS2} other applications based on the spontaneous 
breaking of conformal invariance
and on the scale-invariant extensions of Starobinsky model have been presented.
In Ref.~\cite{cinesi}, a generalization of the Starobinsky model represented by 
a polynomial correction of the Einstein gravity of the type $c_1 R^2+c_2 R^n$ 
has been studied.

In this paper, we will concentrate on inflation caused by scalar curvature 
corrections to Einstein gravity, namely,
we will consider the so-called $F(R)$-gravity, whose action is in the form of 
$F(R)=R+f(R)$, being $f(R)$ a function of the Ricci scalar (for recent reviews 
on modified gravity,
see Refs.~\cite{Review-Nojiri-Odintsov} --\cite{SebRev}  and 
Ref.\cite{others}). This kind of corrections may occur due to quantum effects 
in the
hot universe or maybe motivated by the ultraviolet completation of quantum 
theory of gravity. Our aim is to investigate which kinds of viable inflation 
can be realized in the contest of $F(R)$-gravity beyond the Starobinsky model, 
whose dynamics is governed in the Einstein frame by a potential of the type 
$V(\sigma)\sim (1-\exp[-\sigma])^2$, being $\sigma$ the inflaton.

The paper is organized in the following way. In Sections {\bf 2}--{\bf 3}, we 
will revisit the conformal transformations which permit to pass from the Jordan 
frame to the Einstein one and we will recall the dynamics of the viable 
inflation. In Section {\bf 4}, we will study inflation for the general class of 
scalar potentials of the type $V(\sigma)\sim\exp[n\sigma]$, $n$ being a general 
parameter,
performing the analysis in the Einstein frame and therefore reconstructing the 
$F(R)$-gravity theories which correspond to the given potentials. Viable 
inflation must be consistent with the last Planck data (spectral index, 
tensor-to-scalar ratio...) and must correspond to  Einstein theory corrections 
which emerge only at mass scales larger than the Planck one, namely at high 
curvatures. We will see the conditions on $n$ for which slow-roll conditions 
are satisfied, and we will reconstruct the form of the $F(R)$-models during 
early-time acceleration and their form at small curvaures. Some specific 
examples are presented.
In Section {\bf 5}, following the recent success of
higher derivative gravity, we will revisit and study in detail the specific 
class of models $F(R)=R+(R+R_0)^n$. The analysis in the
Einstein frame reveals that $n$ must be very close to two in order to realize a 
viable inflation for large and negative values of the scalar field,
but other possibilities are allowed
by bounding the field in a different way.
In particular, when $n>2$, the de Sitter solution emerges, but in order to 
study the exit from inflation is necessary to analyze the theory in the Jordan 
frame, where perturbations make possible an early time acceleration with a 
sufficient amount of inflation.
Some summary and outlook are given in Section {\bf 6}.
Technical details and further considerations are presented in Appendixes.

\paragraph*{} We shall use units in which $c=\hbar=k_{\mathrm{B}}=1$,
$c\,,\hbar\,,k_{\mathrm{B}}$  being respectively the speed of light, the
Planck and Boltzmann constants.  Moreover we shall denote by $G_N$
the gravitational constant and  by $M_{\mathrm{Pl}} =G_{N}^{-1/2} =1.2
\times 10^{19}$GeV the Planck mass.
Finally, we shall set   $\kappa^2\equiv 8 \pi G_{N}$.

\section{Conformal transformations}

In scalar-tensor theories of gravity, a scalar field coupled to the metric 
appears in the action.
The first scalar-tensor theory was proposed by Brans \& Dicke in 
1961~\cite{BransDicke} in the attempt to incorporate the Mach's principle into 
the theory of gravity, but today the interest to such theories is related with 
the possibility to reproduce the primordial acceleration of the inflationary 
universe.

In principle, a modified gravity theory can be rewritten in scalar-tensor or 
Einstein frame
form.
Let us start by considering the general action of $F(R)$-modified gravity
\begin{equation}
I = \int_\mathcal{M} d^4 x \sqrt{-g} \left[ \frac{F(R)}{2\kappa^2}\right]\,,
\label{action}
\end{equation}
where $F(R)$ is a function of the Ricci scalar $R$, $g$ is the determinant of 
the metric tensor $g_{\mu\nu}$ and
$\mathcal{M}$ is the space-time manifold.
Now we introduce the field $A$ into
~(\ref{action}),\\
\phantom{line}
\begin{equation}
I_{JF}=\frac{1}{2\kappa^{2}}\int_{\mathcal{M}}\sqrt{-g}\left[
F_A(A) \, (R-A)+F(A)\right] d^{4}x\label{JordanFrame}\,.
\end{equation}
\phantom{line}\\
Here, `$JF$' means `Jordan frame', and $F_A(A)$ denotes the derivative of 
$F(A)$ with respect to $A$.  By making the variation
with respect to $A$, we immediatly obtain $A=R$, such that (\ref{JordanFrame}) 
is equivalent to (\ref{action}) .
We define the scalar field $\sigma$ (which in fact encodes the new degree of 
freedom in the theory, namely the scalaron or inflaton) as
\begin{equation}
\sigma := -\sqrt{\frac{3}{2\kappa^2}}\ln [F_A(A)]\,.\label{sigma}
\end{equation}
By considering the conformal transformation of the metric,
\begin{equation}
\tilde g_{\mu\nu}=\mathrm{e}^{-\sigma}g_{\mu\nu}\label{conforme}\,,
\end{equation}
we finally get the `Einstein frame' action
\phantom{line}
\begin{eqnarray}
I_{EF} &=& \int_{\mathcal{M}} d^4 x \sqrt{-\tilde{g}} \left\{ 
\frac{\tilde{R}}{2\kappa^2} -
\frac{1}{2}\left(\frac{F_{AA}(A)}{F_A(A)}\right)^2
\tilde{g}^{\mu\nu}\partial_\mu A \partial_\nu A - 
\frac{1}{2\kappa^2}\left[\frac{A}{F_A(A)}
- \frac{F(A)}{F_A(A)^2}\right]\right\} \nonumber\\ \nonumber\\
&=&\int_{\mathcal{M}} d^4 x \sqrt{-\tilde{g}} \left( 
\frac{\tilde{R}}{2\kappa^2} -
\frac{1}{2}\tilde{g}^{\mu\nu}
\partial_\mu \sigma \partial_\nu \sigma - 
V(\sigma)\right)\,,\label{EinsteinFrame}
\end{eqnarray}
\phantom{line}\\
where $\tilde{R}$ denotes the Ricci scalar evaluated in the conformal metric 
$\tilde{g}_{\mu\nu}$ and $\tilde{g}$ is the determinant of the conformal 
metric, namely $\tilde{g}=\mathrm{e}^{-4\sigma}g$.
Furthermore, one has\\
\phantom{line}
\begin{equation}
V(\sigma)\equiv\frac{A}{F'(A)} - 
\frac{F(A)}{F'(A)^2}=\frac{1}{2\kappa^2}\left\{\mathrm{e}^{\left(\sqrt{2\kappa^2/3}\right)\sigma}
R\left(\mathrm{e}^{-\left(\sqrt{2\kappa^2/3}\right)\sigma}\right)
-\mathrm{e}^{2\left(\sqrt{2\kappa^2/3}\right) 
\sigma}F\left[R\left(\mathrm{e}^{-\left(\sqrt{2\kappa^2/3}\right)\sigma}\right)\right]
\right\}\,,\label{V}
\end{equation}
\phantom{line}\\
being $R(\mathrm{e}^{-\sqrt{2\kappa^2/3}\sigma})$ the solution of
Eq.~(\ref{sigma}) with $A=R$, namely $R$ a function of 
$\mathrm{e}^{-\sqrt{2\kappa^2/3}\sigma}$.
In what follows, we will omit the tilde to denote all the quantities evaluated 
in the Einstein frame.

Note that string-inspired inflationary models also contain canonical 
and/or tachyon scalar (for recent discussion see Refs.~\cite{Samiinfl}--\cite{Jamil:2013nca}).

\section{Dynamics of inflation}

In this Section, for the sake of completeness, we will recall the well-known 
facts on inflation. The energy density and pressure of the inflaton $\sigma$ 
are given by
\begin{equation}
\rho_\sigma=\frac{\dot\sigma^2}{2}+V(\sigma)\,,\quad 
p_\sigma=\frac{\dot\sigma^2}{2}-V(\sigma)\,,
\end{equation}
where the dot is the derivative with respect to the cosmological time. The 
Friedmann equations in the presence of  $\sigma$ read
\begin{equation}
\frac{3 
H^2}{\kappa^2}=\frac{\dot\sigma^2}{2}+V(\sigma)\,,\quad-\frac{1}{\kappa^2}\left(2\dot 
H+3 H^2\right)=\frac{\dot\sigma^2}{2}-V(\sigma)\,,
\end{equation}
and the energy conservation law coincides with the equation of motion for 
$\sigma$ and reads
\begin{equation}
\ddot\sigma+3H\dot\sigma=-V'(\sigma)\,,
\end{equation}
where the prime denotes the derivative of the potential with respect to 
$\sigma$.
From the Friedmann equations we obtain
\begin{equation}
\dot{H}=-\frac{\kappa^2}{2}\left( \rho_\sigma+ p_\sigma 
\right)=-\frac{\kappa^2}{2} \dot\sigma^2\,.
\end{equation}
On the other hand, the acceleration can be expressed as
\begin{equation}
\frac{\ddot{a}}{a}= H^2+\dot{H}=H^2\left(1-\epsilon \right)\,,
\end{equation}
where we have introduced the ``slow roll'' parameter
\begin{equation}
\epsilon=-\frac{\dot{H}}{H^2}\,.
\end{equation}
This parameter may be expressed as a function of the inflaton as
\begin{equation}
\epsilon=\frac{\kappa^2 \dot\sigma^2}{2 H^2} \,.
\end{equation}
We also have
\begin{equation}
\frac{\ddot{a}}{a}=\frac{\kappa^2}{3}\left(V-\dot\sigma^2 \right)\,.
\end{equation}
Thus, the  condition to have an acceleration is $\epsilon <1$ or $\dot\sigma^2 
< V(\sigma)$. There is another slow roll parameter defined by
\begin{equation}
\eta=-\frac{\ddot{H}}{2H\dot{H}}=\epsilon-\frac{1}{2 \epsilon H}\dot\epsilon 
\,.
\end{equation}
As a function of the inflaton one has
\begin{equation}
\eta=-\frac{\ddot\sigma}{H\dot\sigma} \,.
\end{equation}
For the inflation to occur and persist for a convenient amount of time, a quasi 
de Sitter space is required, namely $\dot H$ has to be very small, and, as a 
result,  also the two slow roll parameters have to be very  small,  and one has
\begin{equation}
\dot\sigma^2\ll V(\sigma)\,,
\end{equation}
namely the kinetic energy of the field has to be small during the inflation. As 
a result, the Friedmann equations  reduce to
\begin{equation}
\frac{3H^2}{\kappa^2}\simeq V(\sigma)\,,
\quad
3H\dot\sigma\simeq -V'(\sigma)\,.\label{EOMs}
\end{equation}
It is easy to show that within this slow roll regime,
  the slow roll parameters may be expressed as function of the inflaton 
potential as
\begin{equation}
\epsilon=\frac{1}{2\kappa^2}\left(\frac{V'(\sigma)}{V(\sigma)}\right)^2\,,\quad 
\eta=\frac{1}{\kappa^2}\left(\frac{V''(\sigma)}{V(\sigma)}\right)\,.\label{slow}
\end{equation}
Inflation ends when $\epsilon\,,|\eta|\sim 1$. A useful quantity which 
describes the amount of inflation is e-foldings number $N$ defined by
\begin{equation}
N\equiv\ln \frac{a_f}{a_i}=\int_{t_i}^{t_e} H dt\simeq 
\kappa^2\int^{\sigma_i}_{\sigma_e}\frac{V(\sigma)}{V'(\sigma)}d\sigma\,,\label{N}
\end{equation}
where the indices $i,f$ are referred to the quantities at the beginning and the 
end of inflation, respectively. The required e-foldings number for inflation is 
at least $N\simeq 60$.
The amplitude of the primordial scalar power spectrum is
\begin{equation}
\Delta_{\mathcal R}^2=\frac{\kappa^4 V}{24\pi^2\epsilon}\,,\label{spectrum}
\end{equation}
and for slow roll inflation the spectral index $n_s$ and the tensor-to-scalar 
ratio are given by
\begin{equation}
n_s=1-6\epsilon+2\eta\,,\quad r=16\epsilon\,.\label{indexes}
\end{equation}
The last Planck data constrain these quantities as
\begin{equation}
n_s=0.9603\pm0.0073\,,\quad r<0.11\,.\label{data}
\end{equation}

\section{Reconstruction of $F(R)$ theory from the scalar potential and analysis 
of the inflation in the Einstein frame}

In this Section, we
will study some classes of scalar potential which produce inflation in the 
Einstein frame. The aim is to generalize the Starobinsky model
by considering different behaviour of the scalar potential as 
$V(\sigma)\sim\exp[n\sqrt{2\kappa^2/3}\sigma]$, where $n$ is the parameter on 
which it depends the dynamics of the inflation (slow roll parameters, spectral 
indices...). This analysis is motivated by the possibility to reconstruct 
suitable $F(R)$ corrections to General Relativity in the corresponding Jordan 
frame. By `suitable' we mean corrections that vanish at mass scales smaller 
than the Planck mass $M_{\text{Pl}}$ and give rise to corrections only in the high 
curvature limits, namely during the inflationary period.
Every scalar potential will be confronted with cosmological data.

In order to reconstruct the $F(R)$-gravity which corresponds to a given 
potential, we may start from Eq.~(\ref{V}). By dividing such equation to 
$\exp\left[2\sqrt{2\kappa^2/3}\right]$, and then by taking the derivative with respect to 
$R$, we get
\begin{equation}
R 
F_R(R)=-2\kappa^2\sqrt{\frac{3}{2\kappa^2}}\frac{d}{d\sigma}\left(\frac{V(\sigma)}{e^{2\left(\sqrt{2\kappa^2/3}\right)\sigma}}\right)\,.\label{start}
\end{equation}
As a result, giving the explicit form of the potential $V(\sigma),$ thanks to 
the relation (\ref{sigma}), we obtain an equation for $F_R(R)$, and therefore 
the $F(R)$-gravity model in the Jordan frame.  In this process, one introduces 
the integration constant, which has to be fixed by
requiring that  Eq.~(\ref{V}) holds true.

\subsection{$V(\sigma)\sim c_0+c_1\exp[\sigma]+c_2\exp[2\sigma]$: $R+ 
R^2+\Lambda $-models\label{4.1}}

Let us start with the following inflationary potential
\begin{equation}
V(\sigma)=\left[c_0+c_1\text{e}^{\sqrt{2\kappa^2/3}\sigma}+c_2\text{e}^{2\sqrt{2\kappa^2/3}\sigma}
\right]\,.\label{cucu}
\end{equation}
This is the simplest example and is the minimal generalization of the 
Starobinsky model.
Equation (\ref{start}) gives
\begin{equation}
2c_0F_R^2+c_1F_R-RF_R=0\,.\label{req}
\end{equation}
Assuming $F_R \neq 0$, one gets
\begin{equation}
F_R(R)=-\frac{c_1}{2c_0}+\frac{R}{2c_0}\,.\label{FF}
\end{equation}
Thus, the corresponding Lagrangian of $F(R)$ gravity is given by
\begin{equation}
F(R)=-\frac{c_1}{2c_0}R+\frac{R^2}{4c_0}+\Lambda\,.\label{Staro2}
\end{equation}
Here, $\Lambda$ is a constant of integration, which can be  determined  by 
equation (\ref{V}). The result is
\begin{equation}
\Lambda=\frac{c_1^2}{4c_0}-c_2\,.\label{L}
\end{equation}
An important remark is in order. In order to have the correct Einstein-Hilbert 
term, we must put $-c_1/(2c_0)=1$. Thus, we have the class 
of modified quadratic models depending on two constants
\begin{equation}
F(R)=R+\frac{R^2}{4c_0}+c_0-c_2\,.\label{Staro3}
\end{equation}
Furthermore, in the specific case  $c_2=0$, one has
  \begin{equation}
F(R)=R+\frac{R^2}{4c_0}+ c_0\,.\label{cogno}
\end{equation}
This is an interesting model, and in the Appendix B, we will study its static 
sperically symmetric solutions.

The other interesting case is the vanishing  of cosmological constant, namely
\begin{equation}
c_2=c_0.\label{L1}
\end{equation}
Since we are assuming $c_1/(2c_0)=1$, it also follows 
$c_2=-c_1/2$. As a consequence,   we recover the
Starobinsky model $F(R)=R+\frac{R^2}{4c_0}$, and related Einstein frame 
potential
\begin{equation}
V(\sigma)=c_0 \left(1-\text{e}^{\sqrt{2\kappa^2/3}\sigma}
\right)^2\,.\label{VStaro}
\end{equation}
The model  (\ref{Staro3}) is the extension of the Starobinsky model to the case 
with cosmological constant and gives a viable inflation.
In what follows, we denote
\begin{equation}
c_0=\frac{\gamma}{4\kappa^2}\,.
\end{equation}
The initial value of the inflaton is large (and negative) and it rolls down 
toward the potential minimum at $\sigma\rightarrow 0^-$, 
$V(0)=-\gamma/(4\kappa^2)<0$. During inflation ($\sigma\rightarrow-\infty$) 
equations (\ref{EOMs}) lead to
\begin{equation}
H^2\simeq\frac{\gamma}{12}\,,
\quad
3H\dot\sigma\simeq 
\left(\frac{\gamma}{\sqrt{6\kappa^2}}\right)\text{e}^{\sqrt{2\kappa^2/3}\sigma}\,.\label{copy}
\end{equation}
It means, that a quasi de Sitter solution can be realized.
We must require
\begin{equation}
\gamma\propto M^2\,,\quad M\ll M_P\,,\label{gammacond}
\end{equation}
where $M_P$ is the Planck mass. As a consequence, the corrections of Eintein
gravity emerge at high curvature and one gets the accelerated expansion with 
$H\propto\sqrt{\gamma}$.
The scalar field behaves as
\begin{equation}
\sigma\simeq 
-\sqrt{\frac{3}{2\kappa^2}}\ln\left[\frac{1}{3}\sqrt{\frac{2\gamma}{3}}(t_0-t)\right]\,,
\end{equation}
where $t_0$ is bounded at the beginning of the inflation. If at this time 
$|\sigma|$ is very large, the slow roll paramters (\ref{slow}),
\begin{equation}
\epsilon=\frac{4}{3}\frac{1}{(2-\text{e}^{-\sqrt{2\kappa^2/3}\sigma})^2}\simeq 
0\,,\quad|\eta|=\frac{4}{3}\frac{1}{|2-\text{e}^{-\sqrt{2\kappa^2/3}\sigma}|}\simeq 
0\,,\label{star}
\end{equation}
are very small and the field moves slowly.
The inflation ends when such parameters are of the order of unit, namely at
$\sigma_e\simeq-0.17\sqrt{3/(2\kappa^2)}$.
The e-foldings number can be evaluated from (\ref{N}) and reads 
($\sigma_i\gg\sigma_e$),
\begin{equation}
N\simeq\frac{3\text{e}^{-\sqrt{2\kappa^2/3}\sigma}}{4}\Big\vert^{\sigma_i}_{\sigma_e}\simeq\frac{1}{4}\sqrt{\frac{2\gamma}{3}}t_0\,.
\end{equation}
The inflation ends at $t_e=t_0-3\sqrt{3/(2\alpha)}\exp[0.17]$. For example, in 
order to obtain $N=60$, we must require 
$\sigma_i\simeq-\sqrt{3/(2\kappa^2)}4.38\simeq 1.07 M_{pl}$. Moreover, we may 
express the e-foldings numbers as
\begin{equation}
\epsilon\simeq\frac{3}{4N^2}\,,\quad|\eta|\simeq\frac{1}{N}\,.
\end{equation}
The amplitude of primordial power spectrum (\ref{spectrum}) is
\begin{equation}
\Delta_{\mathcal R}^2\simeq\frac{\kappa^2\gamma 
N^2}{72\pi^2}\propto\frac{\kappa^2 M^2 N^2}{72\pi^2}\ll \frac{\kappa^2M_P^2 
N^2}{72\pi^2}\,,
\end{equation}
and the indexes (\ref{indexes}) result to be
\begin{equation}
n_s\simeq1-\frac{2}{N}\,,\quad r\simeq\frac{12}{N^2}\,.
\end{equation}
Since  we have $n_s>1-\sqrt{0.11/3}\simeq 0.809$ when $r<0.11\,,n_s<1$, we see 
that these indices are compatible with (\ref{data}). For example, for $N=60$, 
one has $n_s=0.967$ and $r=0.003$.
We stress that this behaviour is the one of Starobinsky model, with a different 
minimum of the potential and a cosmological constant in the Jordan frame. The 
appeareance of a cosmological constant at large curvature needs some 
explanation. It can be originated by some quantum effects or may be supported 
by a modified gravity term which makes it to vanish at small curvatures (see 
for examples the so called `two step-models' in 
Refs.~\cite{twostep1}--\cite{twostep2}). However, if the cosmological constant 
is set equal to zero (or, if necessary, is set equal to an other value), the feature of the model in Einstein frame does not 
change during the inflation: we will see that this double possibility, namely, 
taking a cosmological constant in the Jordan frame, or taking an additional 
term $\propto\exp[2\sqrt{2\kappa^2/3}\sigma]$ in the Einstein frame, does not 
modify the proprieties of the scalar potentials during the early time 
acceleration.

\subsection{$V(\sigma)\sim\gamma\exp[-n\sigma]\,,n>0$: $c_0 
R^{\frac{n+2}{n+1}}$-models\label{case}}

As a second example, we consider the following  potential
\begin{equation}
V(\sigma)=\frac{\alpha}{\kappa^2}\left(1-\text{e}^{\sqrt{2\kappa^2/3}\sigma}\right)+\frac{\gamma}{\kappa^2}\text{e}^{-n\sqrt{2\kappa^2/3}\sigma}\,,
\end{equation}
being $\gamma\,,n>0$ constants. This potential possesses a minimum in which the 
scalar field may fall at the end of inflation. Since for large and negative 
values of the scalar field the potential is not flat, we do not expect a de 
Sitter universe, but if the slow roll conditions are
satisfied, we can obtain an acceleration with a sufficient amount of inflation.

By using our reconstruction,
from (\ref{start}) one derives
\begin{equation}
F_R(R)-\frac{1}{2}+\frac{\gamma}{2\alpha}(2+n) 
F_R(R)^{1+n}=\frac{R}{4\alpha}\,.\label{pippo}
\end{equation}
In principle, this equation admitts many solutions. At the perturbative level, 
it is easy to find that, by putting
\begin{equation}
\alpha=-\gamma(n+2)\,,\label{alphacond}
\end{equation}
we obtain
\begin{equation}
F_R(R\ll\gamma)\simeq 1+c_1 R+c_2 R^2+c_3 R^3+...\,,\label{funzioncina}
\end{equation}
when $|c_1 R|\,,|c_2 R^2|\,, |c_3 R^3|...\ll 1$  (see Appendix A). It means 
that, since 
$c_1\propto\gamma^{-1}\,,c_2\propto\gamma^{-2}\,,c_3\propto\gamma^{-3}...$,
if $\gamma$ satisfies (\ref{gammacond}), we recover the Einstein's gravity when 
$R\ll\gamma$ and the theory is an high curvature correction to General 
Relativity. Moreover, when $R\gg\gamma$, the asymptotic solution of 
Eq.~(\ref{pippo}) with (\ref{alphacond}) is given by
\begin{equation}
F_R(R\gg\gamma)\simeq\left(\frac{1}{4(n+2)}\right)^\frac{1}{1+n}\left(\frac{R}{\gamma}\right)^{\frac{1}{n+1}}\,,\quad
F(R\gg\gamma)\simeq\gamma\left(\frac{n+1}{n+2}\right)\left(\frac{1}{4(n+2)}\right)^\frac{1}{1+n}\left(\frac{R}{\gamma}\right)^{\frac{n+2}{n+1}}\,.\label{expression}
\end{equation}
Here, one important comment is required. In the Einstein frame, 
$R_{EF}\sim\gamma$ during inflation, but the corresponding curvature in the 
Jordan frame is $R_{JF}\simeq \text{e}^{-\sqrt{2\kappa^2/3}\sigma}R_{EF}$ (we 
may neglect the kinetic energy of scalar field in the slow roll approximation), 
such that $R_{JF}\gg\gamma$ when $\sigma\rightarrow-\infty$ and expression 
(\ref{expression}), which is evaluated in the Jordan frame, effectively is 
valid for inflation
(for $n\rightarrow 0$ we recover $F(R)\sim R^2$ in the Jordan frame).

The potential finally reads
\begin{equation}
V(\sigma)=-\frac{\gamma(n+2)}{\kappa^2}\left(1-\text{e}^{\sqrt{2\kappa^2/3}\sigma}\right)+\frac{\gamma}{\kappa^2}\text{e}^{-n\sqrt{2\kappa^2/3}\sigma}\,.
\end{equation}
This potential has a minimum ($V'(\sigma_{min})=0$) at 
$\sigma_{min}=-\sqrt{3/(2\kappa^2)}\log[(n+2)/n]/(n+1)$, and one gets
\begin{equation}
V(\sigma_{min})=\frac{\gamma}{\kappa^2}\left(n^{\frac{1}{n+1}}(n+2)^{\frac{n}{n+1}}+\left(\frac{n+2}{n}\right)^{\frac{1}{n+1}}-(n+2)\right)>0\,,\quad\gamma\,,n>0\,.
\end{equation}
When $\sigma\rightarrow-\infty$ (large curvature), the potential goes to 
infinity, and when $\sigma\rightarrow 0^-$, $V(\sigma)=\gamma/\kappa^2$.
Since the slow roll parameters are given by
\begin{eqnarray}
\epsilon&=&\frac{\left(n-(n+2)\text{e}^{(n+1)\sqrt{2\kappa^2/3}\sigma}\right)^2}{3\left(
(n+2)\text{e}^{n\sqrt{2\kappa^2/3}\sigma}-(n+2)\text{e}^{(n+1)\sqrt{2\kappa^2/3}\sigma}-1
\right)^2}\,,\nonumber\\ \nonumber\\
|\eta|&=&\frac{2}{3}\frac{n^2+(n+2)\text{e}^{(n+1)\sqrt{2\kappa^2/3}\sigma}}{|1+(n+2)\text{e}^{(n+1)\sqrt{2\kappa^2/3}\sigma}-
(n+2)\text{e}^{n\sqrt{2\kappa^2/3}\sigma}
|}\,,
\end{eqnarray}
one has $\epsilon(\sigma\rightarrow-\infty)\simeq n^2/3$ and
$|\eta(\sigma\rightarrow-\infty)|\simeq2n^2/3$, which implies $0<n\ll 1$.
The EOMs (\ref{EOMs})  in the slow roll limit read
\begin{equation}
H^2\simeq\frac{\gamma}{3}\text{e}^{-n\sqrt{2\kappa^2/3}\sigma}\,,\quad
3H\dot\sigma\simeq\sqrt{\frac{2}{3\kappa^2}}\gamma\,n\text{e}^{-n\sqrt{2\kappa^2/3}\sigma}\,.
\end{equation}
The solution for the scalar field is
\begin{equation}
\sigma=\frac{2}{n}\sqrt{\frac{3}{2\kappa^2}}\ln\left[\frac{n^2}{3}\sqrt{\frac{\gamma}{3}}(t_0+t)\right]\,,
\label{SC}
\end{equation}
where $t_0$ is bounded to be very small at the beginning of the inflation, such 
that the field is negative and its magnitude very large. The solution for the 
Hubble parameter finally reads
\begin{equation}
H=\frac{3}{n^2(t_0+t)}\,,\quad\frac{\ddot a}{a}=H^2+\dot 
H=\frac{3}{n^2(t_0+t)^2}\left(\frac{3}{n^2}-1\right)>0\,,\quad (n<\sqrt{3})
\end{equation}
and we have an acceleration as soon as $\epsilon<1$. Despite to the fact that 
in this kind of models the acceleration is smaller than in the de Sitter 
universe, the slow roll paramters can be small enough
to justify our slow roll approximations. A direct evaluation of the ratio of 
kinetic energy  of the field and potential leads to 
$\left(\dot\sigma^2/2\right)/V(\sigma)=n^2/9$, which is much smaller than one 
when $n\ll 1$. The inflation ends when the slow roll paramters are on the order 
of unit, before the minimum of the potential. Note that, by definition, since 
$V'(\sigma_{min})=0$, one has $\epsilon(\sigma_{min})=0$, which corresponds to 
the minimum of the slow roll paramter. However, before to this point, since the 
slow roll parameters behave as
\begin{equation}
\epsilon\simeq\frac{n^2}{3\left(
(n+2)\text{e}^{n\sqrt{2\kappa^2/3}\sigma}-1
\right)^2}\,,\quad
|\eta|\simeq\frac{2}{3}\frac{n^2}{|1-
(n+2)\text{e}^{n\sqrt{2\kappa^2/3}\sigma}
|}\,,
\end{equation}
we find that $\epsilon\,,|\eta|\simeq 1$ when 
$\sigma\simeq-\sqrt{3/(2\kappa^2)}\log[(6+3n)/(3-n\sqrt{3})]/n$.
Finally, from Eq.~(\ref{N}), we get the $N$-foldings number of inflation,
\begin{equation}
N\simeq -\frac{1}{n}\sqrt{\frac{3\kappa^2}{2}}\sigma
\Big\vert^{\sigma_i}_{\sigma_e}\simeq-\frac{3}{n^2}\ln\left[\frac{n^2}{3}\sqrt{\frac{\gamma}{3}}t_0\right]\,.
\end{equation}
When the inflation ends, the field falls in the minimum of the potential and 
starts to oscillate.
The reheating process takes place. The amplitude of primordial power spectrum 
(\ref{spectrum}) and the spectral indexes (\ref{indexes}) can be written as
\begin{equation}
\Delta_{\mathcal R}^2\simeq\frac{\kappa^2\gamma 
\text{e}^{\frac{2}{3}N\,n^2}}{8\pi^2n^2}\,,
\quad
n_s\simeq1-\frac{2n^2}{3}\,,\quad r\simeq\frac{16n^2}{3}\,.
\end{equation}
The corrections to $n_s$ and $r$ are on the order of $\exp\left[-2n^2 
N/3\right]\ll 1$.
These indexes are compatible with (\ref{data}) when
\begin{equation}
n\sim\frac{1}{10}\,,\frac{2}{10}\,.
\end{equation}
These are the typical values of $n$ which make the scalar potential (\ref{V3}) 
able to reproduce a viable inflationary scenario. The result suggests that only 
the models
close to $R^2$-gravity are able to produce this kind of inflation.

In general, $F_R(R)$ in Eq.~(\ref{funzioncina}) may lead to a cosmological 
constant proportional to $\gamma$ in the $F(R)$-model. However, we can set it 
equal to zero, adding a suitable term in
the potential ($\ref{V3}$) proportional to
$\exp\left[2\sqrt{2\kappa^2/3}\sigma\right]$, which
changes much slowler than
$\exp\left[-n\sqrt{2\kappa^2/3}\sigma\right]$ when $0<n$, and the dynamics of 
the inflation is the same of above.

\subsection{$V(\sigma)\sim3\gamma/4-\gamma\exp[\sigma/2]$: $R/2+c_1 R^2+c_2 
(R+R_0)^{3/2}$-models}

We continue our analysis constructing potentials for de Sitter universe during 
inflation,
but with a different behaviour with respect to the Starobinsky one (which 
decreases as $\exp[-\sigma]$). We propose the potential as
\begin{equation}
V(\sigma)=\frac{\alpha}{\kappa^2}-\frac{\gamma}{\kappa^2}\text{e}^{\sqrt{2\kappa^2/3}\sigma/2}\,,\label{V2}
\end{equation}
being $\alpha,\gamma>0$, as usually, constants. It follows
\footnote{Here, we exclude the solution with the minus sign in front of the 
square root which leads to an imaginary value of $\sqrt{F_R}$ when $R$ is 
real.}
\begin{equation}
F_R(R)=\frac{9\gamma^2+8R\alpha+3\sqrt{16 
R\alpha\gamma^2+9\gamma^4}}{32\alpha^2}\,.
\end{equation}
Since we must require $\alpha\,,\gamma\gg 1$ and at small curvature we want to 
recover the Einstein gravity ($F_R=1$), we set
$\alpha=3\gamma/4$, $\gamma>0$, and we get
\begin{equation}
F_R(R)=\frac{1}{2}+\frac{1}{3\gamma}R+\frac{\sqrt{3}}{6}\sqrt{4R/\gamma+3}\,.\label{FFFF}
\end{equation}
Thus, from Eq.~(\ref{V}) we obtain
\begin{equation}
F(R)=\frac{R}{2}+\frac{R^2}{6\gamma}+\frac{\sqrt{3}}{36}\left(4R/\gamma+3\right)^{3/2}+\frac{\gamma}{4}\,.\label{zippozap}
\end{equation}
Here, we stress that the conformal transformation gives for the Ricci scalar
\begin{equation}
R=3\text{e}^{-\sqrt{2\kappa^2/3}\sigma}\left(1+
\text{e}^{\sqrt{2\kappa^2/3/2}\sigma/2}\right)\,,\,
3\text{e}^{-\sqrt{2\kappa^2/3}\sigma}\left(1-
\text{e}^{\sqrt{2\kappa^2/3/2}\sigma/2}\right)\,,
\end{equation}
but only the second one leads to our potential (namely, is the one wich emerges 
from our reconstruction). For $R\ll\gamma$, the model reads $F(R\ll\gamma)\simeq R+\gamma/2$. If we want to recover the General Relativity action $F(R\ll\gamma)\simeq R$, we must set the cosmological constant, namely the last term in (\ref{zippozap}), equal to $-\gamma/4$: in this case
the scalar potential is
\begin{equation}
V(\sigma)=\frac{3\gamma}{4\kappa^2}-\frac{\gamma}{\kappa^2}\text{e}^{\sqrt{2\kappa^2/3}\sigma/2}
-\gamma\frac{\text{e}^{2\sqrt{2\kappa^2/3}\sigma}}{4\kappa^2}\,.\label{V2bis}
\end{equation}
Let us analyze the possibility to reproduce inflation from the potential 
(\ref{V2}), which finally reads
\begin{equation}
V(\sigma)=\left(\frac{3}{4\kappa^2}-\frac{\text{e}^{\sqrt{2\kappa^2/3}\sigma/2}}
{\kappa^2}\right)\gamma\,.
\end{equation}
The initial value of the inflaton is large (and negative) and it rolls down 
toward the potential minimum at $\sigma\rightarrow 0^-$, 
$V(0)=-\gamma/(4\kappa^2)<0$. When $\sigma\rightarrow-\infty$ the EOMs in the 
slow roll limit read
\begin{equation}
H^2\simeq\frac{\gamma}{4}\,,
\quad
3H\dot\sigma\simeq 
\left(\frac{\gamma}{\sqrt{6\kappa^2}}\right)\text{e}^{\sqrt{2\kappa^2/3}\sigma/2}\,.
\end{equation}
It means, that $\gamma$ must satisfy condition (\ref{gammacond}), such that the 
inflation takes place at the Plank epoch.
The de Sitter expansion can be realized and the field behaves as
\begin{equation}
\sigma\simeq 
-\sqrt{\frac{6}{\kappa^2}}\ln\left[\frac{\sqrt{\gamma}}{9}(t_0-t)\right]\,,
\end{equation}
where $t_0$ is bounded at the beginning of the inflation. If at this time the 
magnitude of $\sigma$ is very large, the slow roll parameters (\ref{slow})
\begin{equation}
\epsilon=\frac{4}{3}\frac{1}{\left(4-3\text{e}^{-\sqrt{2\kappa^2/3}\sigma/2}\right)^2}\simeq
0\,,\quad|\eta|=\frac{2}{3}\frac{1}{|4-3\text{e}^{-\sqrt{2\kappa^2/3}\sigma/2}|}\simeq 
0\,,
\end{equation}
are small and the field moves slowly.
The inflation ends at
$\sigma_e\simeq-0.12\sqrt{3/(2\kappa^2)}$, when the slow roll parameters are of 
the order of unit.
The e-foldings number can be evaluated from (\ref{N}) and read,
\begin{equation}
N\simeq\frac{9\text{e}^{-\sqrt{2\kappa^2/3}\sigma/2}}{2}\Big\vert^{\sigma_i}_{\sigma_e}\simeq\frac{\sqrt{\gamma}}{2}t_0\,.
\end{equation}
As a consequence, the slow roll parameters can be written as
\begin{equation}
\epsilon\simeq\frac{3}{N^2}\,,\quad|\eta|\simeq\frac{1}{N}\,.
\end{equation}
The amplitude of primordial power spectrum (\ref{spectrum}) and the spectral 
indexes (\ref{indexes}) are given by
\begin{equation}
\Delta_{\mathcal R}^2\simeq\frac{\kappa^2\gamma N^2}{96\pi^2}\,,
\quad
n_s\simeq1-\frac{1}{N}\,,\quad r\simeq\frac{48}{N^2}\,.
\end{equation}
Since from these formulas we have $n_s>1-\sqrt{0.11/48}\simeq 0.9521$ when 
$r<0.11\,,n_s<1$, we see that these expressions are compatible with 
(\ref{data}). For example, for $N=60$, one has $n_s=0.967$ and $r=0.013$.

We finish this Subsection with some considerations on the potential 
($\ref{V2bis}$), which corresponds to the model with cosmological constant equal to $-\gamma/4$. 
Since the term $\exp\left[2\sqrt{2\kappa^2/3}\sigma\right]$ changes slower than
$\exp\left[\sqrt{2\kappa^2/3}\sigma/2\right]$, the dynamics of inflation is the 
same of above. In this case, when the inflaton exits from the slow roll region, 
it falls in the minimum of the potential located at $V(0)=-\gamma/(8\kappa^2)$.

\subsection{$V(\sigma)\sim\gamma(2-n)/2-\gamma\exp[n\sigma]\,,0<n<1$: $c_1 
R^2+c_2 R^{2-n}$-models}

Now, we would like to investigate some general features of the
inflationary potential
\begin{equation}
V(\sigma)=\frac{\alpha}{\kappa^2}-\frac{\gamma}{\kappa^2}\text{e}^{n\sqrt{2\kappa^2/3}\sigma}\,,\label{V3}
\end{equation}
where $0<\alpha\,,\gamma$ and $0<n<2$. The above potential is explicitly 
constructed to give the de Sitter solution in the slow roll limit when 
$\sigma\rightarrow-\infty$. Equation (\ref{start}) leads to
\begin{equation}
F_R(R)+\frac{\gamma}{2\alpha}F_R(R)^{1-n}(n-2)=\frac{R}{4\alpha}\,.\label{pippo2}
\end{equation}
At the perturbative level, it is easy to see that, by choosing
\begin{equation}
\alpha=\frac{\gamma(2-n)}{2}>0\,,\label{alphaalpha}
\end{equation}
if $\gamma$ satisfies (\ref{gammacond}),
at small curvature one gets
\begin{equation}
F_R(R\ll\gamma)\simeq 1+c_1 R+c_2 R^2+c_3 R^3+...\,,\label{espansione}
\end{equation}
with 
$c_1\propto\gamma^{-1}\,,c_2\propto\gamma^{-2}\,,c_3\propto\gamma^{-3}...$, 
such that  our theory is the high curvature correction to General Relativity 
(see Appendix A).
For example, in the previous Subsection we have seen an exact solution for the 
case $n=1/2$. Since $1\ll \gamma$, when $R/\gamma\ll 1$ we can expand $F_R(R)$ 
in (\ref{FFFF}) as
\begin{equation}
F_R(R\ll\gamma)\simeq 
1+\frac{2}{3\gamma}R-\frac{R^2}{9\gamma^2}+...\,,\label{primeF}
\end{equation}
which returns to be (\ref{espansione}) by using the coefficients in Appendix A.
On the other side, when $R\gg\gamma$, the asymptotic solution of 
Eq.~(\ref{pippo2}) with (\ref{alphaalpha}) is given by
\begin{eqnarray}
F_R(R\gg\gamma)&\simeq&\left(\frac{R}{2\gamma(2-n)}\right)+\left(\frac{R}{2\gamma(2-n)}\right)^{1-n}\,,\nonumber\\\nonumber\\
F(R\gg\gamma)&\simeq&\frac{1}{2}\left(\frac{R^2}{2\gamma(2-n)}\right)+\frac{1}{2-n}\left(\frac{1}{2\gamma(2-n)}\right)^{1-n}R^{2-n}\,.
\label{megastana}
\end{eqnarray}
Also in this case, by taking the exact solution of the prevoius Section in the 
high curvature limit,
\begin{equation}
F_R(R\gg\gamma)\simeq\frac{R}{3\gamma}+\sqrt{\frac{R}{3\gamma}}\,,
\end{equation}
we can verify the consistence of expression (\ref{megastana}) for $n=1/2$.

As an other example, let us consider the case $n=1/3$. The reconstruction leads 
to\\
\phantom{line}\\
\begin{equation}
F_R(R)=\frac{1}{3}+\frac{3}{10}\left(\frac{R}{\gamma}\right)
+\frac{2}{3\times 
5^{1/3}}\left[9\left(\frac{R}{\gamma}\right)+5\right]\frac{1}{\Delta^{1/3}}+\frac{1}{6\times
5^{2/3}}\Delta^{1/3}\,,
\end{equation}
\phantom{line}\\
where
\phantom{line}
\begin{equation}
\Delta=200+243\left(\frac{R}{\gamma}\right)^2+540\left(\frac{R}{\gamma}\right)+27\sqrt{81\left(\frac{R}{\gamma}\right)^4+40\left(\frac{R}{\gamma}\right)^3}\,.
\end{equation}
\phantom{line}\\
At small curvature, it is easy to find
\begin{equation}
F_R(R\ll\gamma)\simeq1+\frac{9}{10}\left(\frac{R}{\gamma}\right)+...\,,
\end{equation}
and in the high curvature limit this model has the following structure,
\begin{equation}
F_R(R\gg\gamma)\simeq 
\frac{3}{10}\left(\frac{R}{\gamma}\right)+\left(\frac{3}{10}\right)^{\frac{2}{3}}\left(\frac{R}{\gamma}\right)^{\frac{2}{3}}\,,
\end{equation}
which corresponds to (\ref{megastana}) with $n=1/3$.
Finally, the model (\ref{FF}) is the limiting case of $n\rightarrow 1$.

Let us analyze this class of potentials. When the magnitude of the inflaton is 
large, the scalar potential (\ref{V3}) with
$\alpha=\gamma(2-n)/2$,
\begin{equation}
V(\sigma)=\frac{\gamma(2-n)}{2\kappa^2}-\frac{\gamma}{\kappa^2}\text{e}^{n\sqrt{2\kappa^2/3}\sigma}\,,\label{n<2}
\end{equation}
behaves as $V(\sigma)\simeq\gamma(2-n)/(2\kappa^2)$, and the EOMs in the slow 
roll limit read
\begin{equation}
H^2\simeq\left(\frac{\gamma(2-n)}{6}\right)\,,
\quad
3H\dot\sigma\simeq 
\left(\frac{n\gamma\sqrt{2}}{\sqrt{3\kappa^2}}\right)\text{e}^{n\sqrt{2\kappa^2/3}\sigma}\,.
\end{equation}
As a consequence, the field results to be
\begin{equation}
\sigma\simeq 
-\sqrt{\frac{3}{2\kappa^2}}\frac{1}{n}\ln\left[\frac{2\sqrt{2}}{3\sqrt{3}}\frac{n^2\sqrt{\gamma}}{\sqrt{2-n}}(t_0-t)\right]\,,
\end{equation}
where $t_0$ is bounded at the beginning of the inflation. When 
$\sigma\rightarrow-\infty$, the slow roll parameters became
\begin{equation}
\epsilon=\frac{4n^2}{3}\frac{1}{\left(2+(n-2)\text{e}^{-n\sqrt{2\kappa^2/3}\sigma}\right)^2}\simeq
0\,,
\quad
|\eta|=\frac{4n^2}{3}\frac{1}{|2+(n-2)\text{e}^{-n\sqrt{2\kappa^2/3}\sigma}|}\simeq 
0\,,
\end{equation}
and are very small.
The inflation ends at
$\sigma_e\simeq-(1/n)\sqrt{3/(2\kappa^2)}\log\left[(\gamma/\alpha)(1-n/\sqrt{3})\right]$,

such that the slow roll parameters are of the order of unit and the field 
reaches the minimum of the potential at $V(0)= -\gamma/(4\kappa^2)$.
The e-foldings number (\ref{N}) is given by
\begin{equation}
N\simeq\frac{3(2-n)\text{e}^{-n\sqrt{2\kappa^2/3}\sigma}}{4n^2}\Big
\vert^{\sigma_i}_{\sigma_e}\simeq
\sqrt{\frac{2(2-n)\gamma}{6}}t_0\,,
\end{equation}
and
\begin{equation}
\epsilon\simeq\frac{3}{4n^2N^2}\,,\quad|\eta|\simeq\frac{1}{N}\,.
\end{equation}
As a consequence, the amplitude of primordial power spectrum and the spectral 
indexes read
\begin{equation}
\Delta_{\mathcal R}^2\simeq\frac{\kappa^2\gamma(2-n)n^2 N^2}{36\pi^2}\,,
\quad
n_s\simeq1-\frac{1}{N}\,,\quad r\simeq\frac{12}{n^2 N^2}\,.
\end{equation}
Since $n_s>1-(\sqrt{0.11/12})n\simeq 1-(0.0957)n$ when $r<0.11\,,n_s<1$, one 
has that
\begin{equation}
0.3386<n<1\,,
\end{equation}
in order to make the spectral indexes compatible with (\ref{indexes}). It 
means, that the potential (\ref{n<2}) can reproduce a viable inflation with at 
least $n=1/3$, namely we are considering models in the form $F(R\gg 
M_{Pl})\simeq c_1 R^2+c_2 R^\zeta$, with $1<\zeta<5/3$.

Also in this case, $F_R(R)$ in Eq.~(\ref{primeF}) may lead to a cosmological 
constant proportional to $\gamma$ in the $F(R)$-model. However, we can set it 
equal to zero, aquiring an additional term in
the potential ($\ref{V3}$) proportional to
$\exp\left[2\sqrt{2\kappa^2/3}\sigma\right]$.
This term
changes slower than
$\exp\left[n\sqrt{2\kappa^2/3}\sigma\right]$ when $0<n<2$, and the dynamics of 
the inflation is the same of above.\\
\\
To conclude this Section, we add some comments about the scalar potentials 
containing exponential terms like $\exp\left[n\sqrt{2\kappa^2/3}\sigma\right]$ 
with $n>1$. Since in this case Eq.~(\ref{alphaalpha}) could lead to $\alpha<0$ 
(when $n>2$) making the potential unable to reproduce inflation, we may 
generalize the potential to the following form
\begin{equation}
V(\sigma)=\frac{\alpha}{\kappa^2}\left(1-
\text{e}^{n\sqrt{2\kappa^2/3}\sigma}\right)
-\frac{\gamma}{\kappa^2}\text{e}^{\sqrt{2\kappa^2/3}\sigma}\,,\label{V4}
\end{equation}
where $\alpha\,,\gamma>0$ and $n>1$.
Now, in order to recover the Einstein gravity at small curvature, we have to 
put
\begin{equation}
\alpha=\frac{\gamma}{2(n-1)}>0\,.
\end{equation}
In the slow roll limit ($\sigma\rightarrow-\infty$) the EOMs read
\begin{equation}
H^2\simeq\frac{\gamma}{6(n-1)}\,,
\quad
3H\dot\sigma\simeq \left(\frac{\sqrt{2}\gamma}{\sqrt{3\kappa^2}}\right)
\text{e}^{\sqrt{2\kappa^2/3}\sigma}\,,
\end{equation}
and the analysis of inflation results to be the same of $R^2$-models (with or 
without the cosmological constant term).\\
\\
Here, we can give some comments about the results of our investigation. Scalar 
inflationary potentials wich satisfy viable conditions for realistic primordial 
acceleration in the contest of large scalar curvature  corrections to General 
Relativity can be classified in two classes. In the first one, the scalar 
potential behaves as 
$V(\sigma)\sim\exp\left[n\sqrt{2\kappa^2/3}\sigma\right]\,,n>0$, and produces 
acceleration with a sufficient amount of inflation only for $n$ very close to 
zero, namely the $F(R)$-model must be close to the Starobinsky one. In the 
second class, $V(\sigma)\sim\alpha 
-\gamma\exp\left[-n\sqrt{2\kappa^2/3}\sigma\right]$, $n>0$, and a quasi de 
Sitter solution emerges during inflation. Cosmological data are satisfied if 
$n>1/3$. This kind of scalar potentials is originated from large scalar 
curvature corrections of the type $F(R)\sim c_1 R^2+c_2 R^\zeta$, with 
$1<\zeta<5/3$ when $1/3<n<1$. Finally, the cases with $1<n$ show in fact the 
same behaviour of $R^2$-model during inflation.

\section{$R+\alpha (R+R_0)^n$-models}

In this Section, we will consider power law corrections to Einstein gravity in 
the form
\begin{equation}
F(R)=R+\alpha (R+R_0)^n+\Lambda\,,\quad n>1\,,
\label{modellino}
\end{equation}
where $\alpha>0$ is a (dimensional) constant parameter and $R_0\,,\Lambda$ are 
two (cosmological) constants introduced to generalize the model. This class of 
models, following the success of quadratic correction, have been often analyzed 
in literature in order to reproduce the dynamics of inflation and
recently, in Ref.~\cite{cinesi}, the cases of polynomial corrections added to 
the Starobinsky model have been investigated.
In what follows, at first we would like to study the
inflation in the Einstein frame given by (\ref{modellino}) with $n\neq2$. We 
will see that inflation for large magnitude values of the field is realized for 
$n\lesssim2$. However, other possibilities are allowed by considering 
intermediate values of the field. In this case, the de Sitter solution may 
emerge in the models with $n>2$, but in order to study the exit from inflation 
is necessary to analyze the theory in the Jordan frame, where perturbations 
make possible an early time acceleration with a sufficient amount of 
$N$-foldings number: that is the aim of the second part of this Section.

Let us start from the potential
  in the scalar field representation of (\ref{modellino}),
\begin{eqnarray}
V(\sigma)&=&\frac{1}{2\kappa^2}\left\{\left(\frac{1}{\alpha 
n}\right)^{\frac{1}{n-1}}\left(1-\text{e}^{\sqrt{(2\kappa^2/3)}\sigma}\right)^{\frac{n}{n-1}}\text{e}^{\frac{(n-2)}{(n-1)}\sqrt{(2\kappa^2/3)}\sigma}\left[\frac{n-1}{n}\right]\right.\nonumber\\
&&\left.+R_0\text{e}^{\sqrt{2\kappa^2/3}\sigma}\left(\text{e}^{\sqrt{2\kappa^2/3}\sigma}-1\right)
-\Lambda\text{e}^{2\sqrt{2\kappa^2/3}\sigma}\right\}\,.
\end{eqnarray}
We immediatly see that only if $n=2$, when $\sigma\rightarrow-\infty$, one gets 
$V(\sigma)\sim\text{const}$ and obtains the de Sitter solution (for $n=2$, 
$\Lambda=R_0=0$ we recover (\ref{VStaro}),
$V(\sigma)=(\exp[\sqrt{2\kappa^2/3}\sigma]-1)^2/(8\alpha\kappa^2)$).
On the other hand, even if $1<n<2$, the scalar field starts from a maximum of 
the potential and falls toward the minimum at $\sigma\rightarrow 0^-$: it 
corresponds to the case of 
$V(\sigma\rightarrow-\infty)\sim\exp[-(2-n)/(n-1)\sigma]$ which may be used to 
reproduce an accelerating expansion if the slow roll limits are satisfied.

The derivatives of the potential read
\begin{eqnarray}
V'(\sigma)&=&\left[-\frac{1}{2\kappa^2}\left(\frac{1}{\alpha 
n}\right)^{\frac{1}{n-1}}
\left(1-\text{e}^{\sqrt{(2\kappa^2/3)}\sigma}\right)^{\frac{1}{n-1}}\text{e}^{\frac{(n-2)}{(n-1)}
\sqrt{(2\kappa^2/3)}\sigma}
+2V(\sigma)\right.
\nonumber\\
&&\left.+\frac{R_0}{2\kappa^2}\text{e}^{\sqrt{2\kappa^2/3}\sigma}\right]\left(\sqrt{\frac{2\kappa^2}{3}}\right)\,,
\end{eqnarray}
\begin{eqnarray}
V''(\sigma)&=&\frac{1}{3}\left(\frac{1}{\alpha 
n}\right)^{\frac{1}{n-1}}\left[\left(\text{e}^{-\sqrt{(2\kappa^2/3)}\sigma}-1\right)^{\frac{2-n}{n-1}}-3\left(\text{e}^{-\sqrt{(2\kappa^2/3)}\sigma}-1\right)^\frac{1}{n-1}\text{e}^{\sqrt{(2\kappa^2/3)}\sigma}\right]
\nonumber\\
&&+4V(\sigma)\left(\frac{2\kappa^2}{3}\right)+R_0\text{e}^{\sqrt{2\kappa^2/3}\sigma}\,,
\end{eqnarray}
and the slow roll parameters (\ref{slow}) are derived as
\begin{equation}
\epsilon=\frac{1}{3}\left[2-\frac{n}{(n-1)}\frac{1}{y}+f(y)\right]^2\,,\quad\eta=\frac{2n}{3(n-1)}\left[\frac{1}{y^2}-\frac{3}{y}\right]+\frac{8}{3}+g(y)\,,
\end{equation}
where we have put
\begin{equation}
y=\left(1-\text{e}^{\sqrt{(\kappa^2/3)}\sigma}\right)\,,\quad
f(y)=\frac{\left[R_0\left(\frac{2n-1}{n-1}\right)-\frac{\Lambda(1-y)}{y}\left(\frac{n}{n-1}\right)\right]}
{\left[\left(\frac{1}{\alpha 
n}\right)^{\frac{1}{n-1}}\frac{y^{\frac{n}{n-1}}}{(1-y)^{\frac{1}{n-1}}}\left(\frac{n-1}{n}\right)+R_0
y-\Lambda(1-y)\right]}\,,\nonumber
\end{equation}
\begin{equation}
g(y)=2f(y)+\frac{\frac{2}{3y}\left(\frac{n}{n-1}\right)\left[\frac{\Lambda(1-y)}{y}-R_0\right]}
{\left[\left(\frac{1}{\alpha 
n}\right)^{\frac{1}{n-1}}\frac{y^{\frac{n}{n-1}}}{(1-y)^{\frac{1}{n-1}}}\left(\frac{n-1}{n}\right)+R_0
y-\Lambda(1-y)\right]}\,.
\end{equation}
We have that $y<1$ for negative values of the field. If $\Lambda=R_0=0$, 
$f(y)=g(y)=0$ and
for $n=2$ we recover (\ref{star}). For large and negative values of the field, 
namely when $y\rightarrow 1^-$, one finds
\begin{equation}
\epsilon\simeq\frac{(2-n)^2}{3(n-1)^2}\,,\quad|\eta|\simeq\frac{4|n-2|}{3(n-1)}\,,
\end{equation}
and $n$ has to be
\begin{equation}
\frac{2+\sqrt{3}}{1+\sqrt{3}}\simeq1.36<n<2\,.\label{nn}
\end{equation}
Here, we remember that we are considering only the cases $n<2$ whose
EOMs in the slow roll limit read,
\begin{equation}
H^2\simeq
\frac{1}{6}\left(\frac{1}{\alpha 
n}\right)^{\frac{1}{n-1}}\text{e}^{\frac{(n-2)}{(n-1)}\sqrt{(2\kappa^2/3)}\sigma}\left[\frac{n-1}{n}\right]
\,,\quad
3H\dot\sigma\simeq
\sqrt{\frac{1}{6\kappa^2}}\left(\frac{1}{\alpha 
n}\right)^{\frac{1}{n-1}}\text{e}^{\frac{(n-2)}{(n-1)}\sqrt{(2\kappa^2/3)}\sigma}\left[\frac{2-n}{n}\right]\,.
\end{equation}
The solution for the field is given by
\begin{equation}
\sigma=2\sqrt{\frac{3}{2\kappa^2}}\left[\frac{n-1}{2-n}\right]\ln\left[
\sqrt{\frac{2}{3n}}\frac{(2-n)^2}{6(n-1)^{3/2}}\left(\frac{1}{n\alpha}\right)^{\frac{1}{2(n-1)}}
(t+t_0)\right]\,.\label{bimbo}
\end{equation}
Since $1<n<2$, $\sigma$ is negative for $t_0>0$ very small bounded at the 
beginning of inflation.
The Hubble parameter reads
\begin{equation}
H=\frac{3(n-1)^2}{(2-n)^2}\frac{1}{(t_0+t)}\,,\quad\frac{\ddot 
a}{a}=\frac{3(n-1)^2((n+1)^2-2)}{(n-2)^4}\frac{1}{(t_0+t)^2}>0\,,
\end{equation}
and we have an acceleration with decreasing Hubble parameter and curvature
\begin{equation}
R=\frac{3(n-1)^2\left(3(n-1)^2-(2-n)\right)}{(2-n)^4(t_0+t)}\,.
\end{equation}
The $N$-foldings number of inflation is given by
\begin{equation}
N\simeq -\frac{(n-1)}{(2-n)}\sqrt{\frac{3\kappa^2}{2}}\sigma
\Big\vert^{\sigma_i}_{\sigma_e}\simeq
-3\left(\frac{n-1}{2-n}\right)^2\ln\left[
\sqrt{\frac{2}{3n}}\frac{(2-n)^2}{6(n-1)^{3/2}}\left(\frac{1}{n\alpha}\right)^{\frac{1}{2(n-1)}}
t_0\right]\,.
\end{equation}
When the inflation ends, the field reaches the minimum of the potential. 
We note that the cosmological constants introduced in the model do not play 
any role at inflationary stage, and change the behaviour of the potential only 
at the end of the inflation.
If 
$\Lambda=R_0=0$, it is easy to see that $V(0)=0$, otherwise the potential 
possesses a minimum before $\sigma=0$ (in the same way of \S~\ref{case}), where 
the field falls and starts to oscillate.

The amplitude of primordial power spectrum (\ref{spectrum}) and the spectral 
indexes (\ref{indexes}) are
\begin{equation}
\Delta_{\mathcal 
R}^2\simeq\frac{(n-1)^3\kappa^2\text{e}^{\frac{2}{3}N\frac{(n-2)^2}{(n-1)^2}}}{16\pi^2(2-n)^2n}\left(\frac{1}{\alpha\,n}\right)^{\frac{1}{n-1}}\,,
\quad
n_s\simeq1-\frac{8(2-n)}{3(n-1)}\,,\quad r\simeq\frac{16(2-n)^2}{3(n-1)^2}\,,
\end{equation}
where we have taken into account that $n$ is close to two. In order to make 
these indexes compatible with (\ref{data}), we must require
$ n\sim 1.8\,, 1.9$,
such that, as a matter of fact, the quadratic corrections with $n=2$ appear to be 
the only possibilities of the type (\ref{modellino}) able to reproduce a viable 
inflation in the Einstein frame. However, it is interesting to extend our 
investigation to the Jordan frame.

\subsection{Inflation in the Jordan frame}

In the first part of this Section, we have seen how model (\ref{modellino}) can 
produce inflation in the Einstein frame representation. We have an early time 
acceleration followed by the end of inflation and the slow roll conditions are 
satisfied if $n$ is very close to two.
For the sake of simplicity, in this Subsection, we 
will put $\Lambda=R_0=0$ in (\ref{modellino}).
Let us return to the Jordan frame.

The Friedmann equation for a generic $F(R)$-model read (in vacuum)
\begin{equation}
\left(R F_R(R)- F(R)\right)-6H^2 F_R(R)-6H\dot F_R(R)=0\,,\label{eq}
\end{equation}
such that for (\ref{modellino}) one derives
\begin{equation}
\alpha(R)^{n-1}\left[R(1-n)\right]-6H^2\left[1+\alpha 
n(R)^{n-1}\right]=6H\alpha n\,(n-1)(R)^{n-2}\dot R\,.\label{First}
\end{equation}
In the high curvature limit (it means, for large and negative values of the 
scalar field $\sigma$ in the Einstein frame) we may consider $\alpha 
n(R)^{n-1}\gg 1$,
\begin{equation}
H^2\simeq\frac{(n-1)}{6n}\left[R-\frac{6n\,H\,\dot R}{R}\right]\,.
\end{equation}
During inflation, the Hubble parameter $H$ evolves slowly and we can consider 
the anologous of the slow roll parameterers which have the same
formal dependence of the Einstein frame, namely we require  $|\dot H|/H^2\ll 
1$, $|\ddot H/(H\dot H)|\ll 1$. Thus, we get
\begin{equation}
\frac{\dot 
H}{H^2}\simeq-\left[\frac{4-2n}{n+4(n-1)^2}\right]\,,\quad\frac{\ddot a}{a}=H^2
+\dot H=1-\frac{4-2n}{n+4(n-1)^2}\,.
\end{equation}
It follows that the expansion is accelerated only if
\begin{equation}
n>\frac{5}{4}=1.25\,.
\end{equation}
This value is close to the one given by the left side of (\ref{nn}). Moreover, 
for $n>2$, we see that $\dot H>0$ and the Hubble parameter, and therefore the 
curvature, grows up with the time and the physics of Standard Model is not 
reached. This result is in agreement with the behaviour of the model discussed 
in the Einstein frame: when $n>2$ and the acceleration emerges in high 
curvature limit, the field cannot reach a minimum of the potential for small 
values and we do not exit from inflation. We could arrive to this conclusion 
also by looking for Eq.~(\ref{bimbo}): when $n>2$, $t_0$ has to be very large 
at the beginning of inflation and the scalar field grows up with the time. 
However, it does not mean that models with $n>2$ do not produce inflation. In 
scalar field representation we start at very large and negative values of the 
field (it means, that in Jordan framework we can ignore the $R$-term in the 
action). It may be interesting to see what happen for intermediate values of 
the field. We analyze the problem in the Jordan frame. From Eq.~(\ref{eq}) we 
have the de Sitter condition at $R_0=12H_0^2$, $H_0$ constant,  namely
\begin{equation}
2F(R_0)-R_0F_R(R_0)=0\,,\label{trace}
\end{equation}
which leads in our case,
\begin{equation}
R_0=\left(\frac{1}{\alpha(n-2)}\right)^{\frac{1}{n-1}}\,.\label{RdS}
\end{equation}
The model with $n=2$ does not possess an exact de Sitter solution, but if 
$n>2$, we can realize it. Here, we remember that $\alpha>0$.
In the Einstein frame this solution corresponds to
\begin{equation}
\sigma_{max}=-\sqrt{\frac{3}{2\kappa^2}}\log\left[\frac{2n-2}{n-2}\right]\,,
\end{equation}
where we have used (\ref{sigma}). We note that $\sigma<0$ only if $n>2$ (we are 
assuming $n>1$); otherwise, if $n<2$, this expression becomes imaginary and 
meaningless.
This solution corresponds to a maximum of the potential, because of
\begin{equation}
V'(\sigma_{max})=0\,,\quad 
V''(\sigma_{max})=\frac{1}{6}\left(\frac{1}{\alpha\,n}\right)^{\frac{1}{n-1}}
\frac{n-2}{n-1}\left(\frac{n}{n-2}\right)^{\frac{(2-n)}{(n-1)}}>0\,.
\end{equation}
When $n>2$, the potential in Einstein frame has a maximum. The scalar field 
produces the de Sitter solution, but the field does not evolve with the time 
($\dot\sigma=0$) and we do not have a natural exit form inflation. However, by 
making use of the perturbative theory, we may study the stability of the model 
in the Jordan frame.

By perturbating Eq.~(\ref{eq}) as $R\rightarrow R+\delta R$, $|\delta R|\ll 1$ 
around the de Sitter solution, we get
\begin{equation}
\left(\Box-m^2\right)\delta R\simeq 0\,,
\quad
m^2=\frac{1}{3}\left(\frac{F'(R)}{F''(R)}-R
\right)\,.
\end{equation}
Here, $m^2$ is the effective mass of the scalaron, namely the new degree of 
freedom
introduced by the modified gravity through $F_R(R)$, which is proportional to 
the opposite of the inflaton, namely $-\sigma$. As a consequence, $m^2$ is 
proportional to the opposite of the scalar potential of the inflaton and when 
$m^2<0$ the solution is unstable. In our case, we get on the de Sitter 
solution,
\begin{equation}
m^2=-(n-2)^{\frac{n-2}{n-1}}\left(\frac{1}{\alpha\,n}\right)^{\frac{1}{n-1}}<0\,,\quad 
2<n\,.\label{inst1}
\end{equation}
It means, that a small perturbation can cause the exit from inflation. Now, the 
question is: in which direction the inflaton moves due to a perturbation and 
which kind of perturbation we need to have a correct $N$-foldings number? To 
reproduce a viable comsology, we expect that the inflaton moves toward the 
minimum of the potential at $V(0)=0$, such that the small curvature regime can 
be reached and the cosmology of Standard Model takes place, and $N$ must be at 
least $N\sim 60$. In principle, the inflaton can also moves to 
$\sigma\rightarrow-\infty$ (large curvature regime), for which the potential 
tends also to zero.

In hot universe scenario,
we must take into account also the presence of
ultrarelativistic matter/radiation, whose energy density is given by
\begin{equation}
\rho_r=\rho_{r(0)}\,a(t)^{-4}\,.
\end{equation}
Here, $\rho_{r(0)}$ is a constant bounded at the beginning of inflation. Thus, 
Eq.~(\ref{First}) reads
\begin{equation}
\frac{3H^2}{\kappa^22}=\rho_{MG}+\rho_r\,,\quad\rho_{MG}=
\frac{1}{2\kappa^2}\left[\alpha(R)^{n-1}\left[R(1-n)\right]-6H^2\alpha 
n(R)^{n-1}-6H\alpha n\,(n-1)(R)^{n-2}\dot R\right]\,,\label{F1}
\end{equation}
where $\rho_{MG}$ encodes the amount of energy density given by the correction 
term $R^n$ to Eintein gravity. For simplicity, we introduce the red shift 
parameter $z=-1+1/a(t)$. We also remember that $d/dt=-(z+1)H(z)d/dz$.

Let us define~\cite{HuSaw, Bamba}
\begin{equation}
y_H (z)\equiv\frac{\rho_{\mathrm{MG}}}{M^2/\kappa^2}=\frac{3H^2}{M^2}
-\tilde\chi (z+1)^{4}\,,\label{yy}
\end{equation}
where $M^2$ is a suitable dimensional constant, such that $[M^2]=[H^2]$ and 
$\tilde\chi=\rho_{r(0)}\kappa^2/M^2$.
By taking into account that the Ricci scalar $R(z)=\left[12H(z)^2-6(z+1)H(z) d 
H(z)/dz\right]$ results to be
\begin{equation}
R=M^2\left(4y_H-(z+1)\frac{d y_H(z)}{dz}\right)\,,
\end{equation}
we derive from (\ref{First}),
\begin{eqnarray}
&&
\frac{d^2y_H(z)}{dz^2}-\frac{dy_H}{dz}\frac{1}{z+1}\left\{3-\frac{f_R(R)}{2M^2f_{RR}(R)\left[y_H(z)+\tilde\chi(z+1)^4\right]}\right\}
\nonumber\\
&&
{}+\frac{y_H(z)}{(z+1)^2}\left\{\frac{1-f_R(R)}{M^2f_{RR}(R)\left[y_H(z)+\tilde\chi(z+1)^4\right]}\right\}
\nonumber\\
&&
{}+\frac{2f_R(R)\tilde\chi(z+1)^4+f(R)/M^2}{(z+1)^2\,2M^2f_{RR}(R)\left[y_H(z)+\tilde\chi(z+1)^4\right]}=0\,,
\label{superEq}
\end{eqnarray}
where in our case
\begin{equation}
f(R)=\alpha R^n\,,\quad f_R(R)=\alpha\,n R^{n-1}\,,\quad 
f_{RR}=\alpha\,n(n-1)R^{n-2}\,.
\label{zazam}
\end{equation}
Let us study the perturbations around the de Sitter solution (\ref{RdS}), 
namely
\begin{equation}
y_{H}(z)\simeq y_{0}+y_1(z)\,,\label{yexpansion}\quad 
y_0=\frac{1}{4M^2}\left(\frac{1}{\alpha(n-2)}\right)^{\frac{1}{n-1}}\,,
\end{equation}
where
$|y_1(z)/y_0|\ll 1$. Here, we have put $R_0=4M^2y_0$ in the de Sitter universe.
By assuming the contribution of ultrarelativistic matter much smaller than 
$y_0$ at the beginning of inflation, Eq.~(\ref{superEq}) becomes, at first 
order in $y_1(z)$,
\begin{equation}
\frac{d^2 y_1(z)}{d z^2}-\frac{2}{(z+1)}\frac{d y_1(z)}{d z}+\frac{1}{(z+1)^2}
\left(-4+\frac{4(1+f_R(R))}{Rf_{RR}(R)}\right)
y_1(z)\simeq0\,,\label{superEqbis}
\end{equation}
where we have also used condition (\ref{trace}). Thus, the solution for $y_1$ 
is given by
\begin{equation}
y_1(z)=C_0(z+1)^x\,,\quad
x=\frac{1}{2}\left(3-\sqrt{25-\frac{16(1+f_R(R))}{R f_{RR}(R)}}\right)\,,
\end{equation}
$C_0$ being constant.
Here, we do not consider the solution with the plus sign in front of the square 
root of $x$, since it does not cause any instability. On the other hand, by 
making use of (\ref{zazam}), one has on the de Sitter solution (\ref{RdS})
\begin{equation}
x=\frac{1}{2}\left(3-\sqrt{\frac{25n^2-57n+32}{n(n-1)}}\right)<0\,,\quad n>2\,,
\end{equation}
since it is easy to demonstrate that $x<0$ if $0<16(n-2)/n$, that is always 
true when $2<n$ and we recover condition (\ref{inst1}).
As a consequence, the perturbation $y_1(z)$ grows up in expanding universe as
\begin{equation}
y_1(z)=y_1(z_\mathrm{i})\left[\frac{(z+1)}{(z_\mathrm{i}+1)}\right]^x\,.\label{yyy}
\end{equation}
Here, we have considered $C_0=y_1(z_\mathrm{i})/(z_\mathrm{i}+1)^x$, $z_i$ 
being the redshift at the beginning of inflation where perturbation is bounded.
When $y_1(z)$ is on the same order of $y_0$, the
inflation ends. A classical perturbation on the (vacuum) de Sitter solution may 
be given by
the ultrarelativistic matter in (\ref{yy}), such that Eq.~(\ref{yyy}) reads
\begin{equation}
y_1(z)=-\tilde\chi(z_i+1)^4\left[\frac{(z+1)}{(z_\mathrm{i}+1)}\right]^x\,,\quad 
y_1(z_i)=-\tilde\chi(z_\text{i}+1)^4\,.
\end{equation}
Thus, the $N$-foldings number during inflation is
\begin{equation}
N=\log \left[\frac{z_\mathrm{i}+1}{z_\mathrm{e}+1}\right]
\simeq \frac{1}{x}\log\left[\frac{\tilde\chi(z_{\text{i}}+1)^4}{y_0}\right]\,.
\label{Nfolding}
\end{equation}
Here, $z_e$ denotes the red shift at the end of inflation and we have 
considered $y_1(z_\text{e})\simeq-y_H$. As small $x$ is, much unstable is 
inflation. It means, that also a small initial perturbation may give arise to a 
large $N$-foldings. We can write $\tilde\chi$ as
\begin{equation}
\tilde\chi=\frac{y_0}{(z_i+1)^4}\times\delta\,,\quad\delta\ll 1\,,
\end{equation}
such that
\begin{equation}
\delta=\text{e}^{x\,N}\,.\label{delta}
\end{equation}
In order to obtain an $N$-foldings of $70$, for $n=3$ ($x\simeq-0.393$) we need 
$\delta\sim10^{-12}$, for $n=4$ ($x\simeq-0.562$) we need $\delta\sim10^{-18}$,
for $n=5$ ($x\simeq-0.656$) we need $\delta\sim10^{-20}$,
for $n=10$ ($x\simeq-0.835$) it is enough $\delta\sim10^{-26}$.

Moreover, the behaviour of Ricci scalar is given by
\begin{equation}
R=4y_0\,M^2+M^2\left(4y_1-(z+1)\frac{dy_1}{dz}\right)\simeq 4y_0M^2-y_0\delta 
M^2
\left[4-x\right]\left(\frac{z+1}{z_\text{i}+1}\right)^x\,.
\end{equation}
Since $x>0$, the Ricci scalar decreases with the red shift and the physics of 
Standard Model can emerge.
In Ref.~\cite{nostrum} numerical calculations have been executed in different 
inflationary models
provided by the specific $R^n$-term with $2<n<3$, which makes inflation 
unstable and
in the presence of ultrarelativistic matter, which leads to the exit from the 
inflationary stage toward the physics of Standard Model.

In conclusion, we can say that large scalar curvature corrections to Einstein 
gravity of the type $R^n$, $n>2$, may represent a valid inflationary scenario. 
In this case, the curvature is bounded at a specific value at the beginning of 
inflation, and the de Sitter space-time is an exact solution of the model, 
which results to be highly unstable. From (\ref{delta}) we see how extremely 
small perturbations in hot universe give the correct $N$-foldings number and 
the exit from inflation.
Finally, it is important to note that positive energy density of perturbations 
brings the curvature to decrease during the early-time acceleration making the 
model consistent with the observable evolution of our universe.

\section{Conclusions}

The attention that recently has been paid to modified theories of gravity is 
caused by the idea  of unified description of early-time and late-time cosmic 
acceleration~\cite{Review-Nojiri-Odintsov}. Moreover,
the gravitational action of such a kind of theories may describe quantum 
effects in hot universe scenario, and the last cosmological data seems to be in 
favour of quadratic corrections to General Relativity during this phase.

In this paper, we have investigated some feautures of $F(R)$-modified gravity 
models for inflationary cosmology, by performing our analysis in the Jordan and 
in the Einstein framework.

At first, we have studied
inflation for the class of scalar potentials of the type 
$V(\sigma)\sim\exp[n\sigma]$, $n$ being a general parameter, in Einstein frame. 
As a matter of fact, for such a kind of models is possible to reconstruct  the 
$F(R)$-gravity theories which correspond to the given potentials.
Since
viable inflation must be consistent with the last Planck data, the potentials 
have been carefully alalyzed, by finding the conditions on the parameters which 
make possible the early-time acceleration according with $N$-foldings, spectral
index and tensor-to-scalar ratio coming from observations.
We have derived the form of the $F(R)$-models at the large and small curvature 
limits, demonstrating that these
models in the Jordan frame correspond to corrections to Einstein gravity which 
emerge only at mass scales larger than the Planck mass.
The investigated potentials can be classified in two classes. In the first one, 
the scalar potential behaves as 
$V(\sigma)\sim\exp\left[n\sqrt{2\kappa^2/3}\sigma\right]\,,n>0$, and produces 
acceleration with a sufficient amount of inflation only for $n$ very close to 
zero, namely the $F(R)$-model must be closed to the Starobinsky one. In the 
second class, $V(\sigma)\sim\alpha 
-\gamma\exp\left[-n\sqrt{2\kappa^2/3}\sigma\right]$, $n>0$, and the quasi de 
Sitter solution emerges during viable inflation when $n>1/3$. This kind of 
scalar potentials is originated from large curvature corrections of the type 
$F(R)\sim c_1 R^2+c_2 R^\zeta$, with $1<\zeta<5/3$ when $1/3<n<1$, or from 
quadratic curvature corrections when $n>1$.
It is interesting to note that $R^2$-term which induces the early-time 
inflation is also responsible for removal of finite-time future 
singularity  in $F(R)$-gravity unifying the inflation with dark energy~\cite{OdSing}.

In the second part of the paper, we have studied in detail the specific class 
of models $F(R)=R+(R+R_0)^n$. The analysis in the
Einstein frame reveals that $n$ must be very close to two in order to realize a 
viable inflation for large and negative values of the scalar field,
but other possibilities are allowed
by starting from intermediate values of the field.
To be specific, when $n>2$, the de Sitter solution emerges, but in order to 
study the exit from inflation, since in this case the de Sitter space-time is 
an exact solution and the field does not move and exit from inflation in a 
natural way, is necessary to analyze the theory in the Jordan frame, where 
perturbations make possible an early time acceleration with a sufficient amount 
of inflation. Moreover, we can explicitly demonstrate
that the curvature decreases making the model consistent with the historical 
evolution of our universe.

\section*{Acknowledgments}
We would like to thank M. Sami for useful discussions and valuable suggestions.
The work by SDO has been supported in part by MINECO (Spain), project 
FIS2010-15640, by AGAUR (Generalitat de Catalunya), contract 2009SGR-994 and by 
project 2.1839.2011 of MES (Russia).


\section*{Appendix A}
Let us consider the equation
\begin{equation}
F_R(R)-\frac{1}{2}-\frac{F_R(R)^{1+n}}{2} =-\frac{R}{4\gamma(n+2)}\,,
\label{EqA0}\end{equation}
with $n\,,\gamma>0$ (it corresponds to (\ref{pippo}) with (\ref{alphacond})).
We want to find solutions of the above equation as a function of $R$.
A simple possibility is looking for regular solutions as a power series of $R$, 
namely
setting 
\begin{equation*}
F_R(R)=1+c_1R+c_2R^2+c_3R^3+...\,,
\end{equation*}
where the constants $c_{1,2,3...}$ have to be determined. At first, we are 
interested in solutions where $|c_{1,2,3...}R^{1,2,3...}|\ll 1$, namely we 
analyze the limit at small curvature. Thus,
plugging this Ansatz in the above equation one has
\begin{equation*}
(1+c_1R+c_2R^2+c_3R^3+...)-\frac{1}{2}-\frac{1}{2}\left(1+c_1R+c_2R^2+ 
c_3R^3+... \right)^{1+n}=
-\frac{R}{4\gamma(n+2)}\,.
\end{equation*}
Putting
\begin{equation*}
X=+c_1R+c_2R^2+ c_3R^3+... \,,
\end{equation*}
and assuming  $|X|\ll 1$, the  following expansion holds
\begin{equation*}
(1+X)^{1+n}=1+(1+n)X+\frac{(1+n)n}{2}X^2+\frac{n(n+1)(n-1)}{3!}X^3+...\,.
\end{equation*}
As a consequence, one arrives at the recursive  relations
\begin{equation*}
c_1=\frac{1}{2\gamma(n-1)(n+2)}\,,\quad
   c_2=-\frac{(1+n)n\,c_1^2}{2(n-1)}\,,\quad
     c_3=-\frac{(1+n)c_1^3}{6}\,, ...\,,
\end{equation*}
from which it follows
\begin{equation*}
c_1 R\propto\frac{R}{\gamma}\,,\quad
    c_2 R^2\propto\frac{R}{\gamma^2}\,,\quad
       c_3 R^3\propto\frac{R^3}{\gamma^3}\,,...
\end{equation*}
  This approximation is valid when $R\ll\gamma$.
On the other side,  when $R\gg\gamma$  we can  check for the solutions in the 
form
\begin{equation*}
F_R(R)=c_0\left(\frac{R}{\gamma}\right)^\zeta\,,\quad\zeta<1\,.
\end{equation*}
In such a case from (\ref{EqA0}) we get
\begin{equation*}
c_0^{1+n}\left(\frac{R}{\gamma}\right)^{\zeta(1+n)}\simeq\frac{1}{4(n+2)}\left(\frac{R}{\gamma}\right)\,,
\end{equation*}
and as a consequence
\begin{equation*}
c_0=\left(\frac{1}{4(n+2)}\right)^{\frac{1}{1+n}}\,,\quad\zeta=\frac{1}{1+n}<1\,.
\end{equation*}
Now let us consider  the following equation (it corresponds to \ref{pippo2} 
with \ref{alphaalpha}),
\begin{equation}
F_R(R)-F_R(R)^{1-n} =\frac{R}{2\gamma(2-n)}\,,
\label{EqA1}\end{equation}
where $\gamma>0$ and $0<n<1$.
The solution for $R\ll\gamma$ is given by
\begin{equation*}
F_R(R)=1+c_1R+c_2R^2+c_3R^3+...\,,
\end{equation*}
with
\begin{equation*}
c_1=\frac{1}{2\gamma\,n(2-n)}\,,\quad
c_2=-\frac{n c_1^2}{2}\,,\quad
c_3=\frac{(1-n)(n+1)c_1^3}{6}\,, ...
\end{equation*}
On the other hand, when $R\gg\gamma$ one can expand $F_R(R)$ as
\begin{equation*}
F_R(R)=\beta_1\left(\frac{R}{\gamma}\right)^{\alpha_1}+
\beta_2 \left(\frac{R}{\gamma}\right)^{\alpha_2}\,,\quad\alpha_1>\alpha_2\,.
\end{equation*}
Since $0<n<1$ at the first order in $R/\gamma$ one finds
\begin{equation*}
\beta_1\left(\frac{R}{\gamma}\right)^{\alpha_1}\simeq\frac{R}{2\gamma(2-n)}\,,
\end{equation*}
and so
\begin{equation*}
\alpha_1=1\,,\quad\beta_1=\left(\frac{1}{2\gamma(2-n)}\right)\,.
\end{equation*}
Moreover, by using (\ref{EqA1}) again  we obtain
\begin{equation*}
\alpha_2=1-n\,,\quad\beta_2=\left(\frac{1}{2\gamma(2-n)}\right)^{1-n}\,.
\end{equation*}
As a final result, we can conclude that  at high curvatures the model
under considertion  reads
\begin{equation*}
F_R(R)=\left(\frac{R}{2\gamma(2-n)}\right)+\left(\frac{R}{2\gamma(2-n)}\right)^{1-n}\,.
\end{equation*}

\section*{Appendix B}
As is well known, the Einstein-Hilbert Lagrangian density modified by a
quadratic term $R^2$ and by  a cosmological constant term has the
Schwarzshild-de Sitter solution,
and this is the most general spherically symmetric, static solution if the
parameters which enter in the modified Lagrangian are arbitrary (unrelated).

Here we shall show that there exists a particular choice of such
parameters for which the model has a more general spherically
symmetric, static solution, which is formally identical to the one
corresponding to the Reissner-Nordstr\"om-de Sitter black hole.

To this aim we consider the Lagrangian density
\begin{equation*}
F(R)=R+\frac{R^2}{4c_0}+c_0\,,
\end{equation*}
which has been obtained from the reconstruction process  described in \S 
\ref{4.1}.
This Lagrangian depends on one free parameter only.

Now it is easy to verify that the most general spherically symmetric,
static solution of field equations in vacuum is given by
\begin{equation*}
ds^2=-A(r)dt^2+ \frac{dr^2}{B(r)}+r^2d\Omega\,,\quad\quad 
A(r)=B(r)=1-\frac{a}{r}-\frac{c_0}{6}r^2+\frac{b}{r^2}\,,
\end{equation*}
where $d\Omega$  is the metric on the two sphere, while
  $a$ and $b$ are arbitrary constants.
The physical interpretation of such two  constants of integration is not an 
easy task.
The presence of the $1/r^2$ term is quite interesting and it will be discussed 
elsewhere.


\end{document}